\documentclass[aps,pra,twocolumn,groupedaddress,showpacs]{revtex4}
\usepackage{graphicx}
\usepackage{dcolumn}
\usepackage{bm}
\begin{document}
\title{
An immunity  against correlation attack on quantum stream cipher by Yuen 2000 protocol
}
\author{Osamu Hirota}%
\email{
hirota@lab.tamagawa.ac.jp
}
\affiliation{
Research Center for Quantum Information Science,
Tamagawa University\\
6-1-1, Tamagawa-gakuen, Machida, Tokyo, 194-8610, JAPAN
}

\author{Kaoru Kurosawa}
\email{
kurosawa@mx.ibaraki.ac.jp
}
\affiliation{
Faculty of Engineering, Ibaraki University\\
Ibaraki, JAPAN
}

\date{\today}
\begin{abstract}
This paper presents the security  analysis on  the quantum stream cipher so called  Yuen-2000  protocol (or $\alpha\eta$ scheme) against the fast correlation attack, the typical attack on  stream ciphers.
Although a very simple  experimental model of the quantum stream cipher without a  random mapper may be decrypted  in the information theoretic sense  by the fast correlation algorithm, it is not a basic feature of Yuen 2000 protocol. 
In fact, we  clarify that there exists a randomization scheme which attains the perfect correlation immunity against such attacks under an approximation. And in this scheme, the running key correlation from the second randomization that determines the mapping patterns is dismissed  also by  quantum noise.
In such a case,  any fast correlation attack does not work on the quantum stream cipher.

\end{abstract}

\maketitle

\section{Introduction}
The quantum stream cipher by quantum noise randomization [1] so called Yuen 2000 protocol might be  an attractive new quantum cryptography which can realize the ultra high speed data encryption.
In fact, many remarkable experiments have been demonstrated [2,3,4].
If this type of the quantum cryptography can provide a pretty good security, it may be widely applied to the real telecommunication networks. So the concrete security analysis on the quantum stream cipher is one of the most interesting subjects. In this paper, we will clarify the security property of the quantum stream cipher against the fast correlation attack, the most typical attack on stream ciphers.

Since the quantum stream cipher by quantum noise randomization has quite different  from the conventional one, we survey the basic structure of stream ciphers.
In the conventional cryptosystem, the stream cipher is implemented by a
pseudo random number generator: PRNG with a short secret key and the XOR operation with plaintext data bits.
Here, let us describe these performances based on an information theoretic approach. The class of stream ciphers is called a non-random stream cipher, when they have the following property:
\begin{equation}
H(Y_n|KX_n) =0
\end{equation}
where $X_n$, $Y_n$ are the $n$-long plaintext and ciphertext, and $K$ is the secret key shared by the legitimate users, respectively. Thus the plaintext and key uniquely determine the
ciphertext. On the other hand, the stream cipher is called the randomized stream cipher when they have the following property:
\begin{equation}
H(Y_n|KX_n) \ne 0.
\end{equation}
This means that ciphertexts are not unique even if the plaintext and key  are same. That is, the ciphertexts are randomized.
In the modern cryptography, this type of cryptography is discussed by  Schnorr, Diffie,
Maurer, and Cachin[5]. Maurer devised a randomized stream cipher for which one can prove that Eve obtains no information in Shannon's sense about the plaintext with probability close to 1. But his protocol works on the assumption that memory capacity of attackers is limited[6].
This approach provides an information theoretic notion of security under a memory restriction. Unfortunately, it is difficult to implement the
practical system with a high speed processing by it. However, it shows that the randomized stream cipher may have a potential of the unbreakable encryption.
Now, Yuen-2000 protocol (Y-00) leads a new type of randomized stream cipher so called "a quantum stream cipher" which can be implemented by the optical communications [1]. A very simple experimental model without a random mapper  may be decrypted in the information theoretic sense by the fast correlation attacks. But a purpose of the demonstration is to verify the quantum noise effect, and it is not a basic feature. 
Y-00 may have a specific  security which cannot be attained by any conventional symmetric key cipher.
In this paper, we will clarify a property of the security of Y-00 against the correlation attacks and show a physical method to attain the perfect correlation immunity against them.

\section{Yuen 2000 protocol}
According to  a quantum detection theory[7],  we have the following properties
for the average error probability:
\begin{equation}
P_e(BP) < P_e(BM), \quad P_e(BP) < P_e(MP)
\end{equation}
where $BP$, $BM$, $MP$ mean a binary pure state, a binary mixed state, and a $M$-ary pure state, respectively. One can apply the above
principle of the quantum detection theory to cryptography.
Y-00 protocol is an example. In Y-00 protocol, a PRNG with a shared key  is used to
make a difference in the performances of quantum signal detection. It means that if Eve does not know the key,  she has to detect $M$-ary quantum states, while Bob's detection procedure is the binary with the key.
So  Eve's detection suffers intrinsic errors.
As a result, Eve has to search for the data or the key based on her detection results with an unavoidable quantum error.

Some ideas for the implementation are proposed as follows[2,3,4].
Alice and Bob share a secret key $K$. The key length is $|K|= 100 \sim 1000$ bits. The key is stretched by a PRNG. The length of the running key is $|K'| \cong 2^{|K|}$. The output bit sequence of the PRNG is divided by each $log M$ bits, and each $log M$ bits is regarded as the running key: $K_R=\{1,2,\dots, K_i, \dots, M\}$. The running key corresponds to the basis $\{|\alpha e^{i\theta_i}\rangle, |\alpha e^{i(\theta_i+\pi)}\rangle \}$. 
That is, when a running key appears, a coherent state basis corresponding to the running key is chosen. Then, the data $x\in X$ is transmitted by $|\alpha e^{i\theta_i}\rangle$, or $ |\alpha e^{i(\theta_i+\pi)}\rangle $ of the basis. 
A mapping pattern  that a mapping function from  running keys to   bases of coherent states is given by the next relation in the basic model of Y-00 by the phase modulation:
\begin{equation}
{\cal{L}}=
\left(
\begin{array}{cccc}
K_i\\
\theta_i 
\end {array}
\right ) 
=
\left(
\begin{array}{ccccc}
1 & 2 & 3 & \dots &M\\
 \theta_1 &\theta_2&\theta_3& \dots&\theta_M
\end {array}
\right ) 
\end{equation}
where the mapping $K_i \rightarrow \theta_i$  means that $K_i \rightarrow \{\theta_i, \theta_i + \pi \}$, and $\pi > \theta_{i+1} > \theta_i > 0$.
However, to employ a random mapping from running keys to   bases of coherent states by an additional LFSR with $K_2$ has been recommended in a real implementation[3]. Here, to simplify the explanation we use Eq(4) as a mapping pattern.
Quantum state sequences emitted from the transmitter can be described as follows:
\begin{eqnarray}
|\Psi \rangle &=&|\alpha(K_{R},X) \rangle_1 |\alpha(K_{R},X) \rangle_2
|\alpha(K_{R},X) \rangle_3 \dots \nonumber \\
&=&|\alpha_i \rangle_1 |\alpha_j \rangle_2 |\alpha_k \rangle_3 \dots
\end{eqnarray}
where $|\alpha_i \rangle$ is one of 2$M$ coherent states, $\alpha_i
=|\alpha|e^{i\theta_i}$, and $i,j,k\in {\cal{M}}=(1 \sim 2M)$.

Alice and Bob will design  the number of basis and signal distance between the neighboring states which satisfy
\begin{equation}
|\langle \alpha_i |\alpha_{i+1} \rangle|^2 \sim 1, 
\end{equation}
If Eve uses a heterodyne measurement as a sub quantum optimum receiver,  Eve's ability can be evaluated by 
\begin{equation}
P_e(i+1| i) = \frac{1}{2} - \frac{1}{\sqrt{2\pi}}\int_{0}^{t_0}\exp(-t^2/2) dt =0.2 \sim 0.5
\end{equation}
where $t_0= \frac{\pi |\alpha|}{2M}$ for the phase modulation scheme, and
$t_0= \frac{|\alpha_{max}- \alpha_{min}|}{4M}$ for amplitude or intensity modulation scheme. This corresponds to the error probability between neighboring states, and gives the degree of the quantum noise effect for the quadrature amplitude $\alpha$.

The output sequence of the transmitter in Y-00  is that sequence of coherent states which convey the information data or key. Even if the sequence of coherent states is a deterministic sequence, Eve has to measure the sequence, and   error in the measured data is inevitable. 
Such an error provides a randomization by quantum noise at the measurement. 
The measured data corresponds to the ciphertext, and the ciphertext is not unique even if the signal is the same one. This fact corresponds to Eq(2).  Despite it, Bob can decrypt the measured data.
Indeed the decision making of legitimate users have no error or few errors because of the measurement with the key, which corresponds to  no encryption. 
Thus a crucial point of Y-00 is to realize the encryption by the unavoidable error of Eve.
So  it is clear that Y-00 is an encryption by quantum noise.

Here we denote the necessities for the evaluation  of the security of Y-00.  In the case of the ciphertext only attacks, we need the following unicity distance:
\begin{equation}
n_{u}=\min\{n:H(K|Y_n)=0\}  \sim \infty
\end{equation}
In the case of the known or chosen plaintext attacks,  the security can be evaluated by the following unicity distance:
\begin{equation}
n_{Gu}=\min\{n:H(K|Y_n X_n)=0\}
\end{equation}
This property is a main subject for the security analysis of Y-00.
There is no general theory on this problem, but we  discuss the feature on the specific attacks.

\section{Concrete model of Attack}
First let us discuss a role of the no cloning property in attacks on Y-00.
If Eve can make many copies of the output state signals by a cloning procedure, she can try the brute force attack on copies of the signals by the receivers with  possible  keys. 
However, in Y-00, the output of the transmitter is the sequence of $2M$-ary coherent states as  the non-orthogonal quantum states for Eve.  Consequently  Eve cannot get the required copies of  the non-orthogonal quantum states according to the no cloning theorem[8,9].
So  Eve cannot launch a parallel optical processing for the attack by all kind of key.
Thus,  Eve cannot realize the parallel processing. But she can make the equivalent situation  to the parallel processing on a long time series sequence  of the coherent state signals.
That is, Eve can try to measure  many different sequential segments of quantum states by the  receivers with many different keys. 
So she can try to decrypt Y-00  along the time axis by ciphertext only attacks with a statistical analysis.
However, if the trial number is $\sim 2^{|K|}$, Eve will spend the intractable time in the real world even if she has the unlimited computation power. This is one of the quantum advantages against the brute force attack. In addition, a quantum unambiguous discrimination attack  also does not work[4].

Consequently, the most important attack is a post measurement procedure which requires 
 a sequential processing on the measurement results along the time axis by a single quantum optimum receiver to discriminate $2M$ coherent states.
In  schemes to get the serial data by the single receiver,  a quantum noise effect is also unavoidable. So Eve cannot get the exact data from her measurement without the key.  This fact realizes a randomized stream cipher as Eq(2).
However, Eve can apply the correlation attacks based on the post measurement procedure, which are  the efficient known plaintext attacks on  stream ciphers [10-13].  Here, we will show a security analysis against the correlation attack based on the quantum individual measurement. 

Let us survey  the
structure of the transmitter of Y-00.
\begin{itemize}
\item[\rm(i)]
The sequence of the output of the PRNG is the sequence of 0 and 1.
\item[\rm(ii)]
The output sequence is divided  by  $log M$ bits, and each block is
transformed into the number of $mod (M)$. These
correspond to  the running key sequence, and assign one basis from $M$ basis sets.
\item[\rm(iii)]
One plaintext is sent by one running key of $log M$ bits.
\end{itemize}
Then, the situation of Eve is described as follows:
\begin{itemize}
\item[\rm(i)]
Eve measures each slot by the heterodyne receiver.
\item[\rm(ii)]
When Eve knows the plaintext, the measurement results are regarded as the running key sequence. Then the measurement results of one slot correspond to
$log M$ bits of the output sequence from the PRNG.
\end{itemize}
So we have to consider the following problem. \\
{\it To determine the structure of the PRNG from the received sequence with errors.}\\
The brute force complexity of this problem is about $O(2^{|K|})$. 
Here, a fast correlation attack is applicable to the above problem when the PRNG is  a linear feedback shift register: LFSR or its nonlinear combinations, only when the secret part of the LFSR is  the initial state.
The basic notion of the fast correlation attacks is to avoid the factor $2^{|K|}$ and derive algorithms with the complexity of order $2^{\eta |K|}$ with respect to $\eta < 1$.
In the correlation attack, the approach of viewing the problem as a decoding problem is used. The linear complexity of the target LFSR  is  $|K|$ and the set of possible LFSR sequences is ${\cal {C}} =2^{|K|}$. For a fixed length $N$ of the measured date, the truncated sequences from $\cal{C}$ form a linear $[N, |K|]$ block code.
The LFSR sequences is regarded as a code word from an $[N,|K|]$ linear block code, and the observed sequence is regarded as the  output of  a binary symmetric channel: BSC which represents the randomness by the non-linear combiner of many LFSRs. Let $P_b$ be an error probability or a crossover probability in the BSC.
Due to the correlation between the input and  output of the channel (BSC), attackers can search for the initial state of the LFSR by the maximum likelihood decoding procedure.
In [11], it is claimed that a fast correlation attack  may succeed  if $N> n_0$, where $n_0$ is the critical length
\begin{equation}
n_0 = \frac{|K|}{C} 
\end{equation}
where $C=1-H(P_b)$ is the channel capacity of the BSC in the fast correlation attack model, and where $H(P_b)=-P_b\log P_b -(1-P_b)\log (1-P_b)$. 
As an example, the complexity of fast correlation attacks is given [11]
\begin{equation}
F\sim O((\frac{1}{2\epsilon})^{t-1} \times 2^{|K|(1-\frac{|K|}{n_0})})
\end{equation}
where $t$ is a number of tap of the  LFSR, and $\epsilon=1/2-P_b$.  
To break the system the attacker needs to observe a segment of length $N$, where
\begin{equation}
N \sim O((\frac{1}{2\epsilon})^{t-1}) > n_0
\end{equation}
In the case of the current stream ciphers, when  $\epsilon > \sim 0.05$, the complexity and the required number of the observation may be  efficiently reduced according to several simulations[10-13].

Let us turn to Y-00. The observations of Eve in Y-00 scheme suffer errors by real noises in the quantum measurement process. When the measurement process is regarded as the channel model with quantum noises,  the fast correlation attack against the LFSR as the driver of the ${\it M}$-ary modulator in Y-00 is applicable, when the tap state of the LFSR is opened. We give, here, a simple example.
The error  in Eve's  phase measurement on a very simple model of Y-00 is described by
\begin{equation}
\theta_i \rightarrow \theta_m =\theta_{i\pm e}, \quad \{e =0,1,2 \}
\end{equation}
where $\theta_m$ is the measurement data when $\theta_i$ is true.
If the mapping pattern is Eq(4) which is a deterministic mapping,  the bit error per each $log M$ bits occurs mainly in the last 3 bits of $log M$ bits. In such a case, if the length of the known plaintext is nearly $|K|$ bits, Eve cannot determine the key even if she has an unlimited power of computer. So, she has to try ciphertext only attacks as the brute force attack on the remained keys along the time axis of the sequences because Eve cannot proceed a parallel processing according to the no cloning property. Consequently it takes  intractable time to try the brute force attack. This fact is one of quantum advantages.
However, it seems to qualify if the length of the known plaintext is sufficiently long. That is, Eve may decrypt Y-00 by using a correlation between each $|K|$ bits block(linear complexity of LFSR) in the measured sequence and a set of LFSR of the number  $2^{\lambda |K|}$, $\lambda < 1$. The correlation comes from the error free bits in the serial segments of $log M$ bits.

On the other hand, when we employ a random mapping from the running keys to bases of coherent states by keyed randomization[3,14], the position of the error bits will be diffused. An example of the concrete random mapping method is given by us[15], which improves the security feature  against the fast correlation attacks.
So, even if Eve can get long known plaintext ($N >> n_0$),  a very simple model of Y-00 with a random mapping technique  is secure against the fast correlation attack  in the sense of the computational complexity, because the computational complexity of the fast correlation attacks, which are known at present, is still exponential for a small $\epsilon$.

Thus, despite that we employ the simple LFSR as the driver to the ${\it M}$-ary modulator, the security of Y-00  against the known plaintext attack is sufficiently protected  in the practical sense by quantum noise effects.
If we  employ a non-linear LFSR by a non-linear combiner (or multiplexed sequence generator)  as the running key generator in Y-00, the required length of the observed sequence for the fast correlation attack on 
the non-linear LFSR itself may be an exponential number even if the system is noiseless.  In this case, the running key sequence of Y-00 corresponds to the output sequence of the channel model of the fast correlation attack. The real observation sequence is the output of the cascaded channels consisting of the channel of the fast correlation attack and the channel of the quantum measurement. Thus, Y-00 is always more secure than the conventional stream cipher owing to the quantum noise effect and the no cloning property as concluded in the references [1,3,4].

\section{CORRELATION IMMUNITY}
In practice, there are many technical limits to give the appropriate quantum noise effects  which  guarantee the security. In the attacks based on heterodyne receiver[1,4], the quantum noise is independent of the signal power and the effect is not so large when the signal energy is large. See Eq(13).  So some mapping mechanisms so called "mapper" are necessary for practical applications of Y-00.

A basic idea of the error enhancement techniques for a provable security has been described in the reference [1] so called the deliberate signal randomization (DSR) which is a method without a shared key. 
Besides, the error performance of Bob is degraded, so one will need an appropriate design. 
Let us introduce keyed randomization[14] as mentioned in the previous section. This is a randomization by an additional LFSR with an additional shared key. This  randomization has  such an advantage that it does not affect the error performance of Bob's receiver.  Although it seems, in general, not to be essential for the ultimate security, however, here we  show that the keyed randomization gives  a method to attain the perfect correlation immunity against the fast correlation attacks under the assumption of no quantum state to quantum state correlation being exploited by Eve.

Let us denote here again that the region of the phase error by the quantum noise is  given only by Eq(13), which is a small effect. Our aim is to enhance this effect by the keyed randomization.
One of the roles of the randomization is to make  error positions in  bit sequences from the LFSR  uniform by the effect of the quantum noise even if the phase error by the quantum noise is small. However, it is not sufficient for our purpose.
In order that the additional keyed randomization gives a further effect on the system,  there exists a condition. That is,  keys of  both the driver and the additional randomization  should be hidden by quantum noise effects. The reason will be explained later.

We will show a method which can attain the perfect correlation immunity. We prepare an additional LFSR which will be used to chose  a mapping  pattern  from many mapping patterns ${\cal {L}}_1, {\cal {L}}_2, {\cal {L}}_3, \dots$. However,  each mapping pattern of the set $\{{\cal {L}}_i\}$ is designed as follows:
\begin{eqnarray}
{\cal{L}}_1&=&
\left(
\begin{array}{cccc}
K_i\\
\theta_i 
\end {array}
\right ) 
=
\left(
\begin{array}{ccccc}
1 & 2 & 3 & \dots &M\\
 \theta_1 &\theta_2&\theta_3& \dots&\theta_M
\end {array}
\right ) \nonumber \\
{\cal{L}}_2&=&
\left(
\begin{array}{ccccc}
2 & 3 &  \dots &M  &1\\
 \theta_1+\delta &\theta_2+\delta & \dots& \theta_{M-1}+\delta  &        \theta_M+\delta
\end {array}
\right )\nonumber \\
{\cal{L}}_3&=&
\left(
\begin{array}{ccccc}
3 & 4 &  \dots &1&2\\
\theta_1+2\delta &\theta_2+2\delta & \dots& \theta_{M-1}+2\delta&\theta_M+2\delta
\end {array}
\right )\nonumber \\
\vdots\nonumber \\
{\cal{L}}_M&=&
\left(
\begin{array}{ccccc}
M & \dots &M-1 \\
\theta_1+(M-1)\delta &\dots&\theta_M+(M-1)\delta
\end {array}
\right )
\end{eqnarray}
where $\delta = |\theta_{i+1} - \theta_i|/M$. The crucial point of this method is the shift permutation in the mapping and the degree of $\delta$.
A mapping pattern is chosen by the random sequence of $log M$ bits from the additional second LFSR. After the selection of the mapping pattern, the first LFSR assigns which basis should be used to transmit the information bit. Since the second LFSR is also shared between Alice and Bob, the error performance of Bob is not degraded.
However, Eve has to discriminate $M\times M$ states which have the phase difference $\delta $. 
The quantum noise in the heterodyne receiver affects several states close to the true phase. That is, the standard deviation of the phase measurement is
\begin{equation}
\sigma =\Delta {\theta_m} > 2M\delta=2|\theta_{i+1} - \theta_i|
\end{equation}
Hence the running keys of the first LFSR and the second LFSR are completely hidden by the quantum noise. This fact is important because the running key correlation from the second LFSR that determines the mapping patterns is dismissed. As a result, it becomes a basis independence. That is, all $log M$ bits per signal slot suffer the error by the quantum noise. 

We here employ a wedge approximation  for the quantum noise effect on the phase space in order to evaluate the quantum noise effect. 
The wedge approximation means the following model: Let us cut a circle on the phase space like a wedge based on the center of the circle,  
$\theta_{i+1}$ and $\theta_i$ on the circle. If the phase difference $|\theta_{i+1}-\theta_i|$ is sufficiently small,  the probability distribution of noise  can be regarded as uniform within the standard deviation and zero outside.
Under this approximation, the symbol error is
\begin{equation}
P_e =1- \frac{1}{M},
\end{equation}
then the bit error is 
\begin{equation}
P_b =\frac{1}{2}(1- \frac{1}{M})
\end{equation}
Thus, in Eq(11) and Eq(12), $\epsilon$ is $\frac{1}{2M} \sim 0, M >> 1$. It leads the fact that the required observation number becomes intractable.
So the correlation immunity is attained. This results is valid for any fast correlation attack under  the approximation that Eve looks at each quantum state independently of the others as in the standard correlation attacks, which does not take into account quantum state to quantum state basis correlation. 

\section{Conclusion}
We have analyzed the security  against the fast correlation attack on  the quantum stream cipher by Yuen 2000 protocol, and we have proposed a scheme which attains the perfect correlation immunity under an approximation.


\end{document}